\documentclass[referee, pdflatex, sn-mathphys]{sn-jnl}

\begin{document}

\title[Article Title]{Direction-dependent Dynamics of Colloidal Particle Pairs and the Stokes-Einstein Relation in Quasi-Two-Dimensional Fluids}

\author[1]{\fnm{Noman Hanif} \sur{Barbhuiya}}\email{barbhuiyanoman@iitgn.ac.in}
\author[2]{\fnm{A. G.} \sur{Yodh}}\email{yodh@physics.upenn.edu}
\author*[1]{\fnm{Chandan K.} \sur{Mishra}}\email{chandan.mishra@iitgn.ac.in}

\affil[1]{\orgdiv{Discipline of Physics}, \orgname{Indian Institute of Technology Gandhinagar}, \orgaddress{\street{Palaj}, \city{Gandhinagar}, \postcode{382055}, \state{Gujarat}, \country{India}}}

\affil[2]{\orgdiv{Department of Physics and Astronomy}, \orgname{University of Pennsylvania}, \orgaddress{\city{Philadelphia}, \postcode{19104}, \state{Pennsylvania}, \country{USA}}}

\abstract{Hydrodynamic interactions are important for diverse fluids especially those with low Reynold’s number such as microbial and particle-laden suspensions, and proteins diffusing in membranes. Unfortunately, while far-field (asymptotic) hydrodynamic interactions are fully understood in two- and three-dimensions, near-field interactions are not, and thus our understanding of motions in dense fluid suspensions is still lacking. In this contribution, we experimentally explore the hydrodynamic correlations between particles in quasi-two-dimensional colloidal fluids in the near-field. Surprisingly, the measured displacement and relaxation of particle pairs in the body frame exhibit \textit{direction-dependent} dynamics that can be connected quantitatively to the measured near-field hydrodynamic interactions. These findings, in turn, suggest a mechanism for how and when hydrodynamics can lead to a breakdown of the ubiquitous Stokes-Einstein relation (SER). We observe this breakdown, and interestingly, we show that the direction-dependent breakdown of the SER is ameliorated along directions where hydrodynamic correlations are smallest. In total, the work uncovers significant ramifications of near-field hydrodynamics on transport and dynamic restructuring of fluids in two-dimensions.}

\maketitle
The investigation of hydrodynamics in fluids at low Reynold numbers has a venerable history and continues to yield surprises \cite{purcell1977life, einstein1956investigations, bain2019dynamic, cui2004anomalous, molaei2021interfacial, shani2014long, leunissen2005ionic, riedel2005self, tateno2019influence, happel2012low, bricard2013emergence, zhang2021effective, son2013bacteria, lauga2016bacterial, vereb2003dynamic, ramadurai2009lateral}. Generally, particle transport in such fluids is influenced by hydrodynamic interactions which span near- and far-field length scales \cite{cui2004anomalous, molaei2021interfacial, shani2014long}, and which depend strongly on spatial confinement (dimension) and fluid boundary conditions \cite{beatus2017two, diamant2005hydrodynamic}. In three dimensions (3D), monopole-like hydrodynamic interactions give rise to drag forces on particles in particle-pairs of the same sign in both longitudinal and transverse directions \cite{beatus2017two, happel2012low}. By contrast, in two-dimensions (2D), the asymptotic far-field hydrodynamic solutions exhibit a dipolar flow profile with \textit{longitudinal drag} and \textit{transverse anti-drag} coupling between particles in particle-pairs \cite{shani2014long, cui2004anomalous}. In addition, as the particle packing fraction increases, near-field drag correlations exhibit oscillatory modulations with respect to particle separation that are in-phase with structural signatures such as the particle pair correlation function \cite{diamant2005correlated}. However, the nature of \textit{transverse} anti-drag coupling in the near-field deviates from the far-field dipolar flow profile and remains largely unexplored; for example, phase differences between transverse and longitudinal correlations could exist and, if so, could have consequences in dense suspensions and confined geometries. Our study of rigidly confined colloidal suspensions in 2D sheds light on these issues, revealing direction-dependent transport of particles in colloid-pairs caused both by the contrast in strength between longitudinal drag and transverse anti-drag, and by the phase difference between the particle-separation-dependent oscillations of longitudinal drag and transverse anti-drag. Moreover, the impact of this anisotropy on the Stokes-Einstein relation is elucidated.

\begin{figure}
\centering
\includegraphics[width=0.85\textwidth]{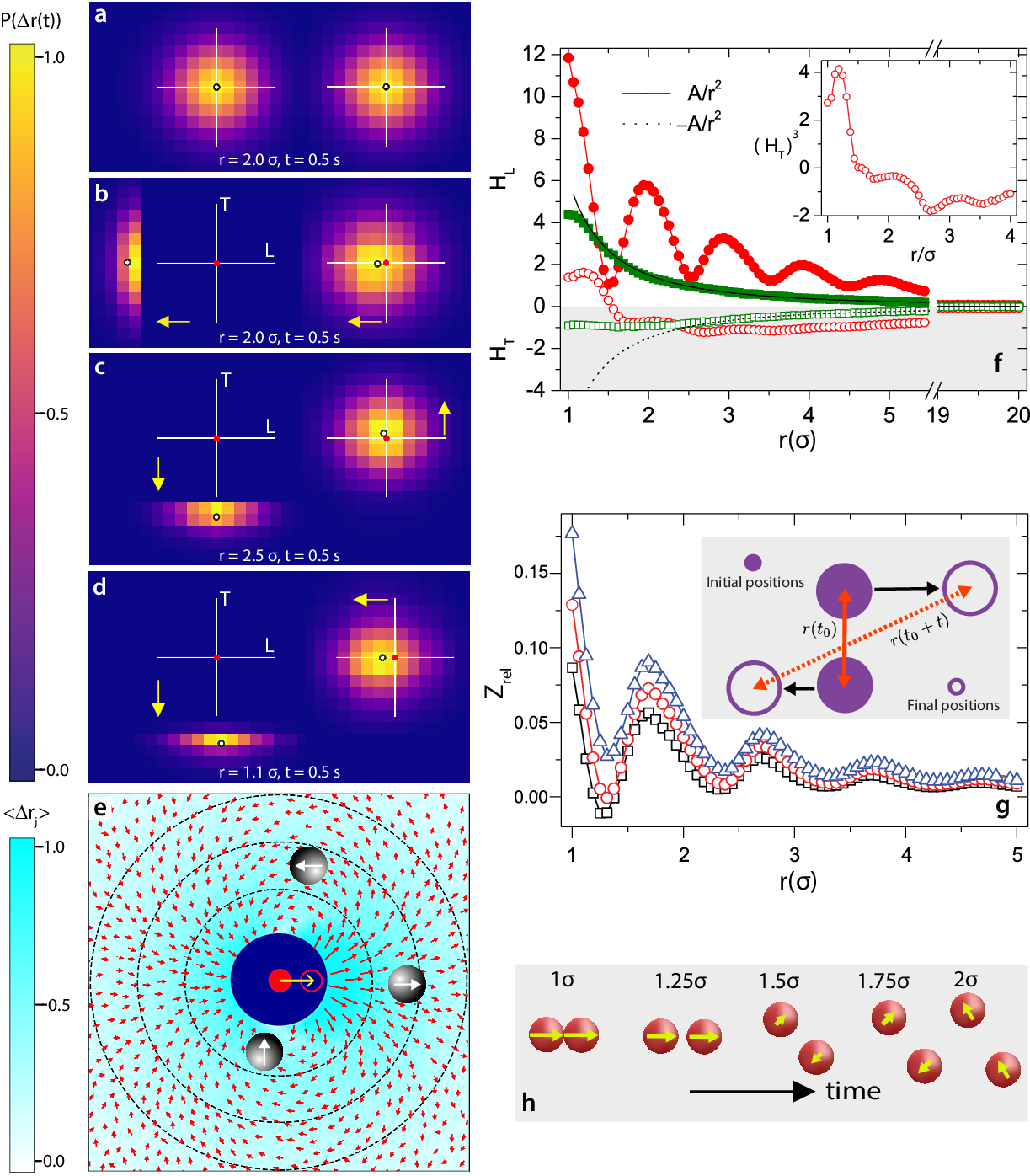}
\caption{$\vline$ \textbf{Visualizing hydrodynamic modes and spatiotemporal evolution of pairs.} \textbf{a}, Colormap of $P(\Delta \textbf{r}_i(\textbf{r}',t) \lvert \textbf{r}_j = \textbf{r}'+\textbf{r})$ and $P(\Delta \textbf{r}_j(\textbf{r}'+\textbf{r},t) \lvert \textbf{r}_i = \textbf{r}')$ for particles in pairs separated by $r = 2.0 \sigma$. Conditional $P(\Delta \textbf{r} (t))$ measured for the particle on the right of the pair when the particle on the left displaces by $ \geq 1.0$ $\sigma$ \textbf{b}, along $L$ and $r = 2.0$ $\sigma$ \textbf{c}, along $T$ and $r = 2.5$ $\sigma$ and \textbf{d}, along $T$ and $r = 1.1$ $\sigma$, depicted by yellow arrows. Solid red and white circles represent the mean positions of the particles at $t_0$ and ($t_0 + t$), respectively. The displacement color map for \textbf{a}$-$\textbf{d} are normalized by the maximum displacements in each case. \textbf{e}, Polar colormap, $ \textbf{r}(r, \theta)$, of hydrodynamic flow profile when the colloid at origin (solid red circle) moves towards the right (open red circle). The dashed radial circles represent $r = \{2, 3, 4\} \sigma$. Representative red arrows, with their head and length represent the direction and strength (shown also in the background), respectively, of the field. The measurements for \textbf{a}$-$\textbf{e} were performed at $\phi = 0.15$ and $t = 0.5$ s. \textbf{f}, $H_L$ (solid symbols) and $H_T$ (open symbols) versus $r$ for $\phi = 0.15$ (green circles) and $\phi = 0.61$ (red squares). The black solid and dashed lines show $\pm A/r^2$ dependencies, respectively, where $A$ is a constant. The inset shows $(H_T)^{3}$ at $\phi=0.61$. \textbf{g}, $Z_{rel}$ versus $r$ at $\phi = 0.61$ for $t = 0.5$ s (black squares), $t = 1.5$ s (red circles) and $t = 3.0$ s (blue triangles). The inset shows typical schematics depicting configurations corresponding to the peak position in $Z_{rel}$. \textbf{h}, ``Most-probable" schematic construction of spatiotemporal evolution of a pair of particles due to near-field hydrodynamics. Note, the relative magnitudes and directions of the yellow arrows correspond to expectation for each configuration.} 
\label{fig1}
\end{figure}

We employ optical video microscopy to probe the hydrodynamic interactions of colloidal suspensions in quasi-two-dimensional (quasi-2D) samples. The experiments were performed with suspensions of micron-size polystyrene latex beads (with diameter, $\sigma$) in their liquid phase as a function of packing area-fraction, $\phi$. Cursory examination of the basic displacement correlation data (Fig. \ref{fig1}) reveals central observations of the experiment. In Fig. \ref{fig1}a, we show the \textit{single particle displacement distribution} after lag time, $t$, for a pair of particles $\{ i, j \}$ located at positions (in 2D) $\{ \textbf{r}', \textbf{r}+\textbf{r}' \}$, respectively; the particles are spatially separated by, $\textbf{r}$, at initial time, $t_0$. The single particle displacement distributions for each particle are $ P(\Delta \textbf{r}_i(\textbf{r}',t) \lvert \textbf{r}_j = \textbf{r}'+\textbf{r})$ and $P(\Delta \textbf{r}_j(\textbf{r}'+\textbf{r},t) \lvert \textbf{r}_i = \textbf{r}')$. Here, displacement distributions are obtained from all $t_0$ and for all pairs of particles separated by distance $r$ at $t_0$. As expected, since particle velocity distributions (and positions) are isotropic, these distributions are measured to be spatially symmetric about their initial position at $t_0$ regardless of $r$ and packing fraction, $\phi$.  

The nature and influence of the hydrodynamic interactions between particles in a colloid-pair are revealed in other panels of Fig. \ref{fig1} (and in other figures) starting with measurements of the \textit{conditional} probability distribution for particle displacements: $P(\Delta \textbf{r}_j(\textbf{r}'+\textbf{r},t) \lvert \Delta \textbf{r}_i^{L, T} (\textbf{r}',t))$. Here $L$ and $T$ denote the longitudinal and transverse axes of the particle pair in the body frame, oriented respectively, along and perpendicular to the line joining the particles separated by $\textbf{r}$ at $t_0$.  $P(\Delta \textbf{r}_j(\textbf{r}'+\textbf{r},t) \lvert \Delta \textbf{r}_i^{L, T} (\textbf{r}',t))$ represents the probability that the $j^{th}$ particle will experience displacement $\Delta \textbf{r}_j(t)$, when the $i^{th}$ particle, separated by $\textbf{r}$, is displaced by $\Delta \textbf{r}_i(t)$ along the longitudinal ($L$) or transverse ($T$) direction in the body frame. Figure \ref{fig1}b-d show example data associated with the conditional distribution of particle displacements. Here, the $i^{th}$ particle is displaced by $\geq \sigma$ during a time lag of $t = 0.5$ s, either along $L$ (Fig. \ref{fig1}b) or along $T$ (Fig. \ref{fig1}c \& d), and the distribution of the $j^{th}$ particle displacement is shown. These conditional probability distributions provide exemplar exhibits of the well-known hydrodynamic dipolar interaction modes \cite{cui2004anomalous, shani2014long}, showing co-diffusion (drag) of particles in a pair along $L$ (Fig. \ref{fig1}b \& SI Video S1) and anti-symmetric diffusion (anti-drag) of particles in a pair along $T$ (Fig. \ref{fig1}c \& SI Video S2). 

Notice, in Fig. \ref{fig1}d, the linear superposition of near-field drag and anti-drag hydrodynamic fields produces circumferential motion of one particle around the other, thereby leading to a ``mass void filling" motion of one particle created by the motion of its partner (Fig. \ref{fig1}d \& SI Video S3). To the best of our knowledge, the full character of this type of mass void filling motion (Fig. \ref{fig1}d) has not been directly observed in experiment. The motional effects caused by near-field drag and anti-drag (and their linear combination) are apparent in the polar plot of the hydrodynamic field profile, which we derive using the ensemble-averaged displacement correlations of particles in a pair with separation vector, $\textbf{r}(r,\theta)$, at $t_0$ and at low $\phi = 0.15$ (Methods, Fig. \ref{fig1}e).

Figure \ref{fig1} also shows the measured longitudinal and transverse displacement correlation functions, $H_L$ and $H_T$, respectively, versus $r$. The longitudinal (transverse) hydrodynamic correlation, $H_{L(T)} (r, t) = {\langle \Delta \textbf{r}_i^{L(T)}(\textbf{r}^\prime, t) \cdot {\Delta \textbf{r}_j^{L(T)}(\textbf{r}^\prime+\textbf{r}, t)} \rangle}/D^{self}$ \cite{shani2014long, cui2004anomalous}. Here, $\Delta \textbf{r}_i^L$ ($\Delta \textbf{r}_i^T$) is the displacement in lag time, $t$, of the $i^{th}-$particle in the $\{i,j\}-$pair along the $L$ ($T$) direction; the averaging, $\langle \rangle$, is performed over all initial times, $t_0$, and all possible unique pairs $\{i,j\}$. Normalization of $H_{L,T}$ by the $\phi-$dependent single-particle diffusivity, $D^{self}$, facilitates comparison of $H_{L,T}$ across different $\phi$ (Methods). Figure \ref{fig1}f shows $H_L$ and $H_T$ versus $r$ for two different packing fractions. At low $\phi$ ($\phi = 0.15$), $H_L$ and $H_T$ exhibit expected dipolar decay profiles in the far-field, \textit{i.e.}, they decay as $1/r^2$. A distinguishing feature of quasi-2D fluid confinement is the positive and negative value of the correlation function amplitude of $H_L$ and $H_T$, respectively; when we remove one cell wall and thereby increase sample dimensionality to 3D, the amplitudes of both $H_L$ and $H_T$ become positive (SI Fig. S1). At higher density, $\phi = 0.61$, local structural features emerge in the near-field (SI Fig. S2). Specifically, oscillatory spatial modulation of the amplitude appears in both $H_L$ and $H_T$, and the hydrodynamic correlation functions deviate from the dipolar form. Nevertheless, in the far-field ($r > 8 \sigma$) even at large packing fraction, the profiles decay as $1/r^2$ (Figure \ref{fig1}f).

As reported in previous studies, our measurements find that the spatial modulation of $H_L$ (in dense suspensions) is in-phase with oscillation of the colloidal fluid structural pair correlation function (SI Fig. S2) \cite{diamant2005correlated, cui2004anomalous}. Surprisingly, we find that the anti-drag spatial modulations associated with $H_T$ exhibit a spatial phase-shift (phase difference/lag) of around $0.25 \sigma$ with respect to $H_L$; this behaviour is most easily seen in the inset to Fig. \ref{fig1}f which plots the cube of $H_T$. This effect is also revealed by the function $Z_{rel} (r,t) \equiv {{\big \langle {r(t+t_0) \over r(t_0)} \big \rangle}_{r^\prime, t_0}-1}$ (Fig. \ref{fig1}g). $Z_{rel} (r,t)$ represents the fractional change in separation of particles in the colloid-pair during lag time $t$. $Z_{rel}$ clearly captures the anti-drag influence on pair-rotation and colloid-pair separation. At the highest $\phi$ ($\phi = 0.61$), $Z_{rel}$ shows oscillatory decaying modulations that are in-phase with the modulations of $H_T$ in the near-field (Fig. \ref{fig1}g).

The insights offered by $H_L$ and $H_T$ (and $Z_{rel}$) suggest a ``most-probable" spatiotemporal evolution of particles in a colloid-pair as a function of the particle separation, $r(t)$. This evolution is schematically shown in Figure \ref{fig1}h. Initially, the particles in the pair are separated by a small distance, $r  \sim 1.0 \sigma$; they then diffuse and separate to $r \sim 1.25\sigma$. When $r \sim 1.25 \sigma$, the pairs are in their most stable state, \textit{i.e.}, they reside in first minima of $Z_{rel}$ (Fig. \ref{fig1}g), and longitudinal drag is dominant. When the separation between particles increases further to $r \sim 1.5 \sigma$, then transverse rotation of the particles begins and leads to further radial separation. When $Z_{rel}$ reaches a maximum at $r \sim 1.75 \sigma$, $H_T$ is comparatively stronger and the pair configuration destabilizes. Over time, as $r$ increases, anti-drag weakens, and drag becomes dominant again at $r \sim 2.0 \sigma$. The cycle will then repeat, but the hydrodynamic interactions become attenuated at larger $r$ (Fig. \ref{fig1}f \& g).

The emergent spatiotemporal mobility landscape, revealed by our experiments in the near-field, leads to particular local viscosity and diffusivity associated with the motions of particles in the colloid-pair. The experiments thus offer an opportunity to explore the validity of basic physics rules such as the Stokes-Einstein relation (SER) in quasi-2D. Recall, the SER relates the particle diffusion coefficient, $D$, to the viscosity, $\eta$, of the suspending fluid: $D = {k_BT \over {6 \pi \eta (\sigma/2)}}$, where $k_BT$ is the thermal energy. In practice, the structural relaxation time, $\tau_\alpha$, is often used as a proxy for $\eta$ \cite{hodgdon1993stokes, larini2008universal, sengupta2013breakdown}. Simulations of $D$ and $\tau_\alpha$ in 3D liquids demonstrate $D \propto \tau_\alpha^{-\xi}$, with expected SER exponent $\xi = 1$ \cite{sengupta2013breakdown}. However, recent computer simulations and experiments in 2D fluids have observed $\xi > 1$ \cite{mishra2015shape, sengupta2013breakdown, perera1998origin}. The origin of this unusual behaviour, which apparently violates the SER, is unresolved. One interesting suggestion alludes to the presence of long-wavelength Mermin-Wagner fluctuations in 2D liquids \cite{li2019long}. These correlations due to Mermin-Wagner fluctuations can be removed by considering the relative motion of particles with respect to their cages \cite{vivek2017long, illing2017mermin}, which, after implementation, recovers $\xi \sim 1$ \cite{li2019long}, and thereby suggests that Mermin-Wagner fluctuations cause the anomalous SER exponent. However, strictly speaking, this approach to filter out correlated motions necessarily assumes that $D$ and $\tau_\alpha$ are isotropic, \textit{i.e.}, the approach assumes that near-field dynamics have zero angular dependence.

\begin{figure}
\centering
\includegraphics[width=1\textwidth]{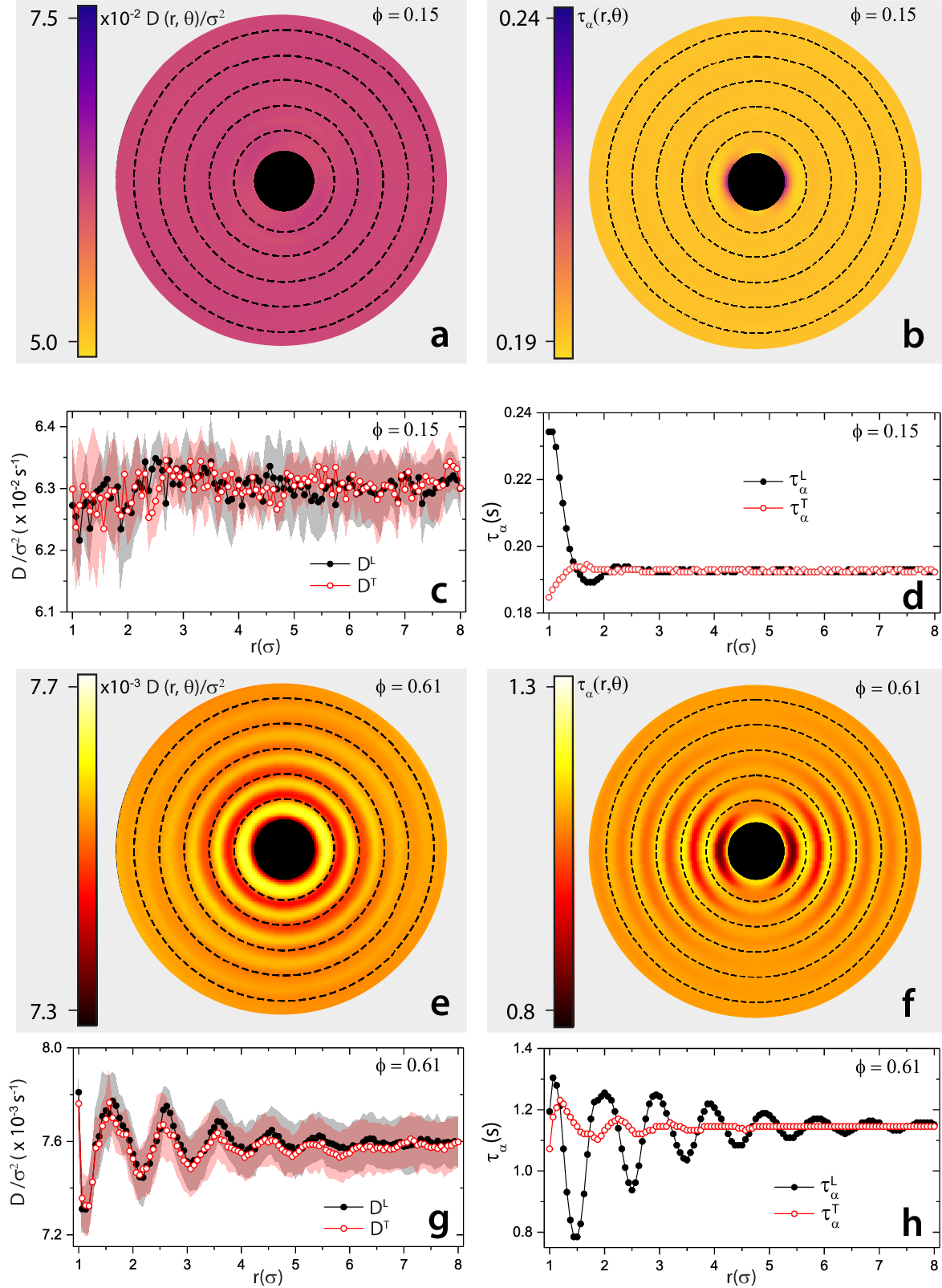}
\caption{$\vline$ \textbf{Elucidating the influence of hydrodynamics on transport quantifiers.} Polar colormaps versus $r$ with color-scale on left for $D(r, \theta)$ at \textbf{a}, $\phi = 0.15$ and \textbf{e}, $\phi = 0.61$, and $\tau_\alpha(r, \theta)$ at \textbf{b}, $\phi = 0.15$ and \textbf{f}, $\phi = 0.61$. The dashed radial circles are at $r = \{2, 3, 4, ...\} \sigma$. Plots along $L$ and $T$ directions corresponding to $\theta = 0^\circ$ and $\theta = 90^\circ$, respectively, for $D(r)$ at \textbf{c}, $\phi = 0.15$ and \textbf{g}, $\phi = 0.61$, and $\tau_\alpha(r)$ at \textbf{d}, $\phi = 0.15$ and \textbf{h}, $\phi = 0.61$. The error bars in $D$ are from fittings.}
\label{fig2}
\end{figure}

Since we have measured longitudinal and transverse hydrodynamic correlations and particle displacements in the near- and far-field, our experiments offer means to revisit the SER in 2D and to directly investigate the influence of spatial phase differences between longitudinal and transverse hydrodynamic modes. Specifically, we study the separation and angular dependence associated with colloid-pair dynamics based on measurements of single particle diffusion, $D(r,\theta)$, and relaxation, $\tau_\alpha (r,\theta)$ (Methods). Here, $r$ is the particle separation distance in a pair at $t_0$, and $\theta$ is the angle between probing direction and the longitudinal axis in the body frame, $L$ (Methods). At low $\phi$ ($\phi = 0.15$), $D(r,\theta)$ is measured to be isotropic (Fig. \ref{fig2}a \& c), but $\tau_\alpha (r,\theta)$ is not; for $r < 2.0 \sigma$, $\tau_\alpha (r,\theta)$ is found to be anisotropic (Fig. \ref{fig2}b \& d). This $\theta-$dependence is readily understood. Since drag leads to co-diffusion of particles in colloid-pairs along the longitudinal direction, when $r \sim 1.25 \sigma$, $H_T$ is substantially smaller than $H_L$, and particles in the colloid-pairs will take longer to relax along $L$ than along $T$: $\tau_\alpha^L (r \sim 1.25 \sigma) > \tau_\alpha^T (r \sim 1.25 \sigma)$ (Fig. \ref{fig2}b \& d). The oscillatory structural features in the near-field become more pronounced when the particle packing area-fraction is increased. $D(r,\theta)$ = $D(r)$ is still measured to be angularly isotropic and exhibits oscillations as a function of $r$ that are in-phase with $H_L$ (Fig. \ref{fig2}e \& g), but $\tau_\alpha (r,\theta)$ is measured to be oscillatory with $r$ and strongly anisotropic (Fig. \ref{fig2}f \& h) due to contrasting magnitude of $H_L$ and $H_T$, and the spatial $\sim 0.25 \sigma$ phase-lag of $H_T$ with respect to $H_L$.  This behaviour is readily apparent in Figure \ref{fig2}d \& h which shows the different variations of $\tau_\alpha^L$ and $\tau_\alpha^T$. 

\begin{figure}
\centering
\includegraphics[width=1\textwidth]{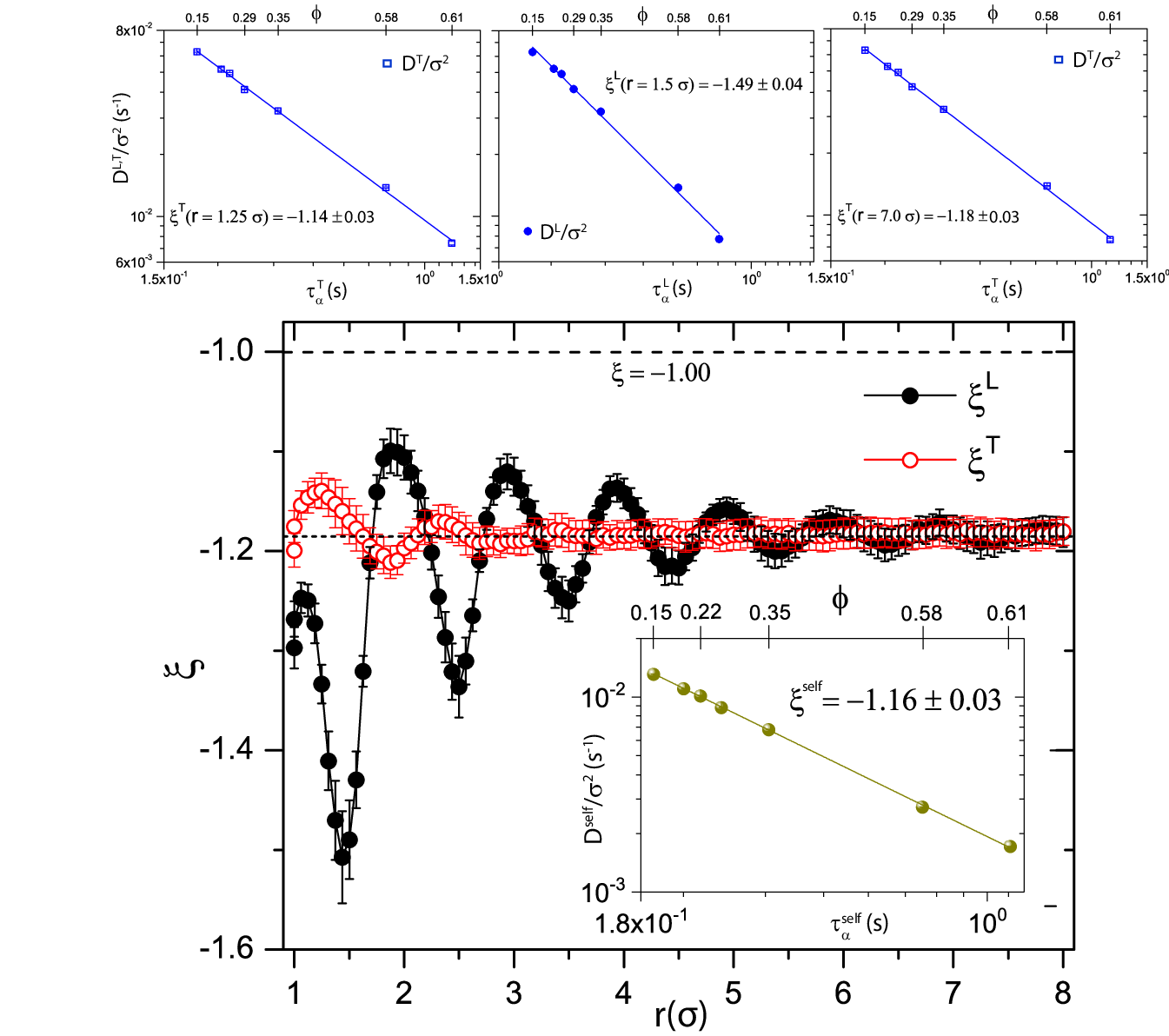}
\caption{$\vline$ \textbf{Influence of hydrodynamics on Stokes-Einstein relation in near-field.} SE exponents $\xi^L$ and $\xi^T$ versus $r$. The inset plots the particle self-diffusivity (derived from measurements in the lab frame), $D^{self}$, versus relaxation, $\tau_\alpha^{self}$; the solid line shows $D^{self} \propto \tau_\alpha^{-1.16 \pm 0.03}$. Black dashed and dotted lines at $\xi = -1.00$ and $\xi = -1.18$ depict the ideally expected and measured asymptotic values of $\xi$, respectively. Top panel shows representative $D^L$ and $D^T$ versus $\tau_\alpha^L$ and $\tau_\alpha^T$, respectively, for different $r$ as shown in the figures. The solid lines depict linear fits to determine $\xi$. Standard error from power-law fittings between $D^{L,T}$ and $\tau_\alpha^{L,T}$ are used in $\xi^{L,T}$ versus $r$ plots; systematic errors, obtained by extraction of $D^{L,T}$ from different time-windows, are found to be larger than standard error and are used when quoting the value of $\xi^{L,T}$ in the main text.}
\label{fig3}
\end{figure}

To explore influence of anisotropy in the hydrodynamic correlations on the validity of the SER, we measured $D(r,\theta)$ and $\tau_\alpha(r, \theta)$ for all packing fractions, $\phi$. For simplicity, our discussion will focus on data taken along the longitudinal ($L, \theta = 0^\circ$) and transverse ($T, \theta = 90^\circ$) directions as a function of $\phi$ for fixed $r$. Specifically, for each $r$, we measure the power-law relationship between $D^{L,T}$ and $\tau_\alpha^{L,T}$ using all $\phi$. Exemplar plots and extracted exponents $\xi^{L,T}$ are shown in the top panel of Figure \ref{fig3}. The resultant variations of $D$ and $\tau_\alpha$ with $r$, and the anisotropy in $\tau_\alpha$ along $L$ and $T$ are reflected in the SER exponents, $\xi^L$ and $\xi^T$, respectively (Fig. \ref{fig3}). Notably, the SER exponents associated with the spatial directions $L$ and $T$ differ from unity and differ from each other. Moreover, the spatial phase lag of $\sim 0.25 \sigma$ observed for $H_L$ and $H_T$ is also apparent in $\xi^L$ and $\xi^T$. By contrast, if instead we derive $\xi$ from measurements of $D$ and $\tau_\alpha$ along two randomly chosen orthogonal directions in the lab frame (different from $L$ and $T$ in the body frame), then we find that the $\xi$ are essentially in-phase and identical within experimental certainty (SI Fig. S3). Together, these observations suggest that the unusual trends of $\xi(r, \theta)$ that are apparent in the body frame are due to the distinct motional modes associated with near-field hydrodynamic correlations that arise in 2D colloidal fluids.  

The consequences of the near- and far-field hydrodynamic correlations persist in our measurements of the traditional single particle diffusivity, $D^{self}$, and the traditional fluid structural relaxation, $\tau_\alpha^{self}$. The $\phi-$dependent $D^{self}$ and $\tau_\alpha^{self}$ yield $\xi^{self} = 1.16 \pm 0.03$ (Inset to Fig. \ref{fig3}). Notice, the spatial modulations of the body-frame $\xi$ decay with $r$ and converge to $\xi^{self}$ in the far-field, $r > 8 \sigma$ ($\xi^{L,T} \rightarrow 1.18 \pm 0.03$). Evidently, the hydrodynamic interactions in quasi-2D that lead to direction-dependent dynamics and SER violation in the body frame, also lead to violation of the SER in the lab frame.  

\begin{figure}
\centering
\includegraphics[width=0.85\textwidth]{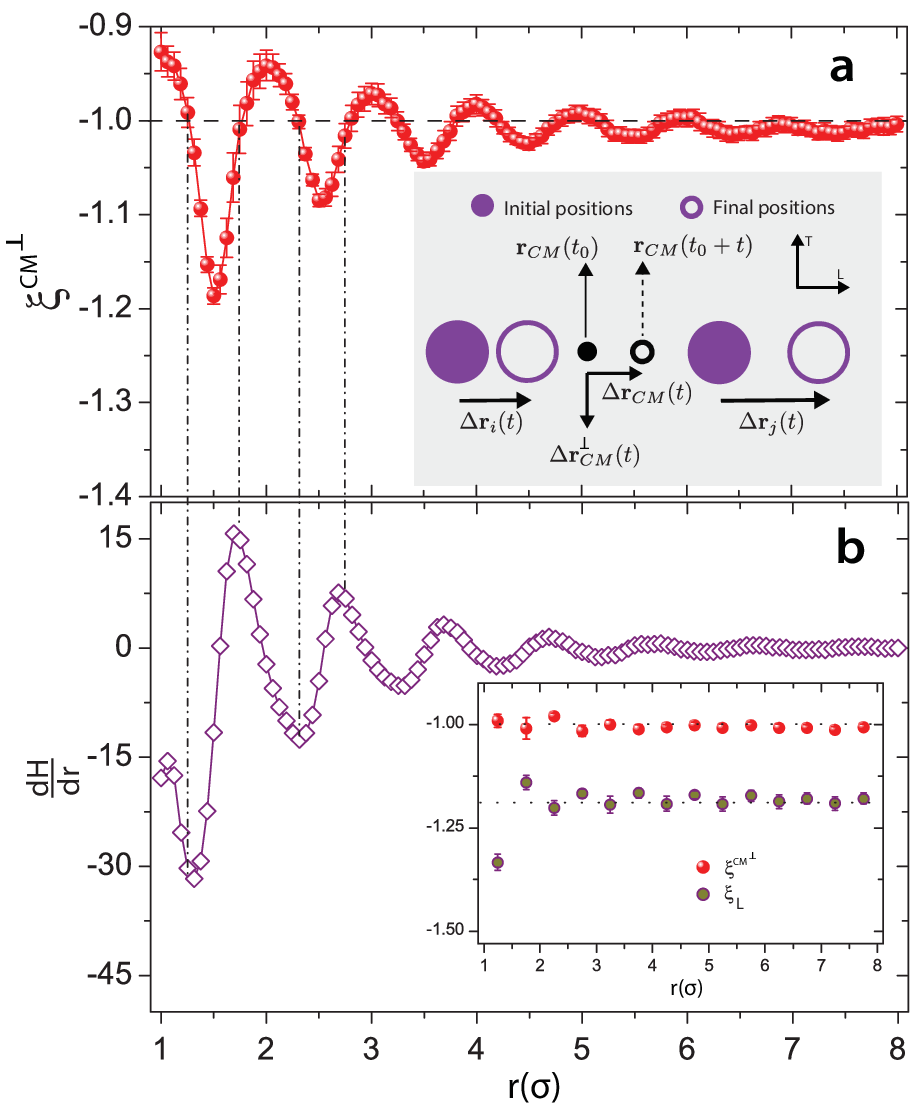}
\caption{$\vline$ \textbf{Recovery of the expected Stokes-Einstein exponent.} \textbf{a}, $\xi^{CM^\perp}$ versus $r$. $\xi^{CM^\perp} \rightarrow -1.01 \pm 0.02$ for $ r>8\sigma$; the ideal expected value of $\xi$ is shown by black dashed line. Standard error from power-law fittings between $D^{CM^\perp}$ and $\tau_\alpha^{CM^\perp}$ are used in $\xi^{CM^\perp}$ versus $r$ plot; systematic errors, obtained by extraction of $D^{CM^\perp}$ from different time-windows, are found to be larger than standard error and are used when quoting the value of $\xi^{CM^\perp}$ in the main text. Inset: schematic to visualize the direction in which the displacements of particles in the pair are least correlated, \textit{i.e.}, direction perpendicular to the centre-of-mass displacement direction. \textbf{b}, $\frac{dH}{dr}$ versus $r$ at $\phi = 0.61$; $H = H_L + H_T$. Inset: comparison of $\xi_L$ with $\xi^{CM^\perp}$ at discrete values of $r$ corresponding to the extrema of $\frac{dH}{dr}$.}
\label{fig4}
\end{figure}

Finally, we consider whether it might be possible to recover the $\xi \sim 1$ SER exponent for quasi-2D colloidal fluids, perhaps along special directions. To this end, we propose a simple approach again based on the colloid-pairs and their correlated interactions and displacements. Generally, validity of the SER ($\xi = 1$) is expected for purely random processes. Since hydrodynamic correlations are smallest along the direction perpendicular to the centre-of-mass displacement ($CM^{\perp}$) of the colloid-pair (inset to Fig. \ref{fig4}a), one might expect that extraction of $D^{CM^{\perp}}$ and $\tau_\alpha^{CM^{\perp}}$ along this direction could yield $\xi^{CM^{\perp}} \sim 1$. The data in Figure \ref{fig4}a corroborates these hypotheses. In the far-field, $r> 8 \sigma$, where the spatial modulations in $H_L$ and $H_T$ are diminished (Fig. \ref{fig1}f), we find that $\xi^{CM^{\perp}}$ decays and saturates to $1.01 \pm 0.02$ (Fig. \ref{fig4}a). In the near field, $r < 8\sigma$, $\xi^{CM^{\perp}}$ oscillates around unity. Interestingly, in this regime ($r < 8\sigma$), $\xi^{CM^{\perp}} \rightarrow 1$ at specific $r$ that correspond to the extrema of $\frac{dH}{dr}$(or extrema of $Z_{rel}$) wherein the net hydrodynamic correlations are weakest in direction orthogonal to the centre-of-mass displacements (Fig. \ref{fig4}b). In the inset to Figure \ref{fig4}b, $\xi^{CM^{\perp}}$ at extrema of $\frac{dH}{dr}$ are compared to the corresponding value of $\xi^L$ in the near-field. Evidently, thermal forces are the dominant fluctuations experienced by particles in the direction orthogonal to the centre-of-mass displacements, and thus the SER is recovered. 

To conclude, we have measured and studied the near- and far-field longitudinal and transverse hydrodynamic modes in quasi-2D colloidal fluids. The findings highlight the importance of the contrasting magnitudes and phase-shift between these modes. These intrinsic features of 2D spatially confined systems lead to spatially inhomogeneous and anisotropic correlated dynamics of particles in colloid-pairs, and to the breakdown of the Stokes-Einstein relation (SER exponent $\xi > 1$). The microscopic insights gleaned suggest a mechanistic route to understand the unusual magnitude of $\xi$ observed here and in other studies \cite{mishra2015shape, li2019long}. Looking forward, these insights about near-field hydrodynamics of spherical particles could prove even more interesting for anisotropic particles, both passive and active, and may lead to novel ideas for affecting self-assembly and structural relaxation \cite{singh2016universal, zhang2021effective, witten2020review, chepizhko2022resonant}. Broadly, we expect that these near-field hydrodynamics could impact phenomena in dense, spatially constrained systems such as arise in cluster aggregation \cite{ginot2018aggregation}, translocation of proteins \cite{vereb2003dynamic, ramadurai2009lateral}, nucleation and growth kinetics of crystals \cite{leunissen2005ionic, tateno2019influence}, active systems \cite{bricard2013emergence, zhang2021effective}, and clogging and jamming of channels \cite{wyss2006mechanism, hong2017clogging}.

\section*{Methods}
\noindent \textbf{Experimental details.} We used polystyrene microspheres of diameter $2 \sigma = 1.04$ $\mu$m with polydispersity of $< 5\%$ suspended in water. The particles were loaded into a wedge-shaped cell and allowed to sediment under gravity into the thin quasi-two-dimensional (quasi-2D) region of the cell. Once a desired packing area-fraction, $\phi$, was achieved, the cell was equilibrated for at least six hours before performing video microscopy. Data for all $\phi$s were taken from the same region of the cell. The images, at each $\phi$, were captured at 10 frames per second (fps) for 20 minutes. The trajectories of the particles were obtained using standard tracking algorithms \cite{crocker1996methods}. The dynamic spatial resolution was found to be $20$ nm. All subsequent analyses were performed using in-house developed codes.

\noindent \textbf{Hydrodynamic flow profile.} The hydrodynamic field shown in \ref{fig1}e was determined as follows \cite{molaei2021interfacial}. Briefly, we first computed the displacements of each particle, $\Delta \textbf{r}_i(t = 0.5 \text{s})$. Next consider the pair of particles, $\{i, j\}$. $\Delta \textbf{r}_i(t = 0.5 \text{s})$, $\Delta \textbf{r}_j(t = 0.5 \text{s})$, and the unit vector ($\hat{\textbf{r}}$) pointing from $i$ to $j$ subtend angles $\alpha_i$, $\alpha_j$, and $\beta_{ij}$ with respect to the positive $x-$axis (ranging between $0$ and $2\pi$). For each reference particle $i$ we rotate the 2D coordinate system through an angle $\alpha_i$ so that $\Delta \textbf{r}_i(t = 0.5 \text{s})$ is aligned in the positive $x$ (horizontal) direction. The other vectors then also rotate through $\alpha_i$. We define the $\theta$ as the polar angle between the positive $x-$axis (now aligned with $\Delta \textbf{r}_i(t = 0.5 \text{s})$) and $\hat{\textbf{r}}$ (SI Fig. S4). The polar plot ($r$, $\theta$) in \ref{fig1}e is derived from the ensemble and initial time, $t_0$, average of the displacements.

\noindent \textbf{Dynamics measurement.} Single particle diffusivity, $D^{self}$, in the lab frame were measured from the mean squared displacements, $\langle \Delta r(t)^2 \rangle$ (SI Fig. S5). $\langle \Delta r(t)^2 \rangle = \langle \frac{1}{N} \sum_{k=1}^{N} (\Delta \textbf{r}_k(t))^2 \rangle$. Here, $N$ is the total number of particles, $\Delta \textbf{r}_k(t)$ is the displacement of $k^\text{th}$ particle during the lag time, $t$, and the averaging, $\langle \rangle$, were performed over $t_0$. 

The diffusion of single particles, $D (r, \theta)$, with respect to the colloid-pair body-frame were measured from dynamical quantities, $\Delta r^2(t; \textbf{r}', \textbf{r}+\textbf{r}', \theta)$, obtained using displacement of either of the particle $\{i,j\}$ in a pair along $\hat{\textbf{R}}(\theta) \hat{\textbf{r}}(t_0, \textbf{r}', \textbf{r}+\textbf{r}'))$; here $\hat{\textbf{r}}$ is the unit vector along the line joining the particles in pair located at $\{\textbf{r}', \textbf{r}+\textbf{r}'\}$ and at initial time $t_0$, and $\hat{\textbf{R}}(\theta)$ is the rotation matrix. $D$ is obtained from the linear regime of $\langle \Delta r(t)^2 \rangle$ plot. Note, even $D(r)$ at low $\phi$, $\phi \leq 0.35$, are anisotropic when extracted from duration timescales ($t < 20$ s) where hydrodynamic interactions are significant (SI Fig. S6). At higher $\phi$ ($\phi \geq 0.58$), the dynamics become mildly sub-diffusive on short timescales, precluding extraction of $D(r, \theta)$, and hence, we cannot comment on whether $D(r)$ continues to be anisotropic at these densities at short timescales.

The structural relaxation time, $\tau_\alpha^{self}$, in the lab frame were measured from self-intermediate scattering functions, $F_s(\textbf{q}, t)$ \cite{mishra2015shape} (SI Fig. S5). $F_s(\textbf{q}, t) = \langle \frac{1}{N} \sum_{k=1}^{N} e^{i\textbf{q}.\Delta \textbf{r}_k(t)} \rangle$, where symbols have usual meanings as explained above. For all the analyses presented in this study (including pair dynamics), the magnitude of probe wave-vector, $q = {{2\pi}/ a}$, where $a$ is the position of the first peak in the pair correlation function, $g(r)$, at $\phi = 0.61$. The direction of $\textbf{q}$ is chosen to be along $x-$axis. 

The structural relaxation time, $\tau_\alpha (r, \theta)$, associated with particle motion with respect to the colloid-pair body-frame were measured using $F_s(\textbf{q}_\theta, t; \textbf{r}', \textbf{r}+\textbf{r}')$, by using $\Delta \textbf{r}(t; \textbf{r}', \textbf{r}+\textbf{r}', \theta)$ of either of the particle $\{i,j\}$ in a pair along $\hat{\textbf{R}}(\theta) \hat{\textbf{r}}(t_0, \textbf{r}', \textbf{r}+\textbf{r}'))$. $\textbf{q}_\theta$ is along $\hat{\textbf{R}}(\theta) \hat{\textbf{r}}$. The time for which the decay of $F_s(q,t)$ drops to $1/e$ is read-off as structural relaxation time, $\tau_\alpha$, \textit{i.e.}, $F_s(q, t = \tau_\alpha) = 1/e$.

\bmhead{Supplementary information}
This article contains supplementary files/videos.  

\bmhead{Acknowledgments}
Authors thank Rajesh Ganapathy, Prasanna Venkatesh B., Adhip Agarwala, K. Hima Nagamanasa, and Sankalp Nambiar for useful discussions. We gratefully acknowledge financial support from the  Department of Science and Technology (Government of India), INSPIRE fellowship (NHB), Indian Institute of Technology Gandhinagar, India (CKM) and the Start-up Research Grant through SRG/2021/001077, of Science and Engineering Research Board of Government of India (CKM), US National Science Foundation through Grant DMR2003659 (AGY) and the MRSEC Grant DMR1720530 including its optical microscopy shared experimental facility (AGY).

\section*{Declarations}
\begin{itemize}
\item Authors declare no competing financial interests. 
\item Correspondence and requests of materials should be addressed to C.K.M.
\end{itemize}

\bibliography{sn-bibliography}

\end{document}